\documentclass[prb,onecolumn,12pt]{revtex4}
\usepackage{epsfig}
\usepackage{amssymb}
\usepackage{threeparttable}
\usepackage{graphicx}
\usepackage{natbib}
\usepackage{longtable}
\usepackage{txfonts}
\usepackage{color}

\begin{document}
\title{Electronic, transport and optical properties of monolayer $\alpha$ and $\beta-$GeSe: A first-principles study}
\author{Yuanfeng Xu$^1$, Hao Zhang$^{1,4,\dag}$, Hezhu Shao$^2$, Gang Ni$^1$, Hongliang Lu$^3$,  Rongjun Zhang$^1$, Bo Peng$^1$, Yongyuan Zhu$^4$ and Heyuan Zhu$^{1,\ddag}$}
\affiliation{$^1$Department of Optical Science and Engineering and Key Laboratory of Micro and Nano Photonic Structures (Ministry of Education), Fudan University, Shanghai 200433, China\\
$^2$Ningbo Institute of Materials Technology and Engineering, Chinese Academy of Sciences, Ningbo 315201, China\\
$^3$State Key Laboratory of ASIC and System, Institute of Advanced Nanodevices,School of Microelectronics, Fudan University, Shanghai 200433, China\\
$^4$Nanjing University, National Laboratory of Solid State Microstructure, Nanjing
210093, China}
\email{zhangh@fudan.edu.cn;hyzhu@fudan.edu.cn}

\begin{abstract}
The extraordinary properties and the novel applications of black phosphorene induce the research interest on the monolayer group-IV monochalcogenides. Here using the first-principles calculations, we systematically investigate the electronic, transport and optical properties of monolayer $\alpha-$ and $\beta-$GeSe, the latter of which was recently experimentally realized. We found that, monolayer $\alpha-$GeSe is a semiconductor with direct band gap of 1.6 eV, and $\beta-$GeSe displays indirect semiconductor with the gap of 2.47 eV, respectively. For monolayer $\beta-$GeSe, the electronic/hole transport is anisotropic with an extremely high electron mobility of 7.84 $\times10^4$$cm^2/V\cdot {s}$ along the zigzag direction, comparable to that of black phosphorene. Furthermore, for $\beta-$GeSe, robust band gaps nearly disregarding the applied tensile strain along the zigzag direction is observed. Both monolayer $\alpha-$ and $\beta-$GeSe exhibit anisotropic optical absorption in the visible spectrum.
\end{abstract}

\maketitle
\section{INTRODUCTION}
The atomic-layered materials with weak interlayer van der Waals bonding, also called as two-dimensional (2D) materials, have received tremendous attentions since the experimental realization of graphene\cite{sciencegraphene, Novo2005, zhang2005}. Nowadays, 2D materials have formed a large material family, involving various kinds of layered structures and chemical elements, e.g. graphene\cite{alandin2011, Balandin2008, Schedin2007}, transition metal dichalcogenides (TMDs)\cite{Kaasbjerg2012, Conley2013}, stanene\cite{Zhu2015,SMLL201570033}, MX$_2$ (M=Mo, W; X=S, Se)\cite{Yun2012a, Kumar2012}, penta-graphene\cite{Zhang2015,xu2017-1}, and etc.

Recently, black phosphorene, a monolayer composed of phosphorus atoms with puckered structure, attracts much attention for its extraordinary properties, including a direct semiconductor with a moderate band gap of 1.5 eV \cite{Qiao2014, hong2014polarized}, a high hole mobility comparable to graphene\cite{Qiao2014}, strong anisotropic transport property\cite{Han2014, Liu2014, hong2014polarized, Eswaraiah2016} and etc, which make it a promising candidate for future electronic and optoeletronic applications. Many efforts have been devoted on the discovery of new 2D materials with ``phosphorene analogues'' puckered structure\cite{Gomes2015}, since the successful prediction of black phosphorene. The monolayer of group-IV monochalcogenides MX(M=Ge, Sn; X=S, Se), which was reported to be thermoelectric materials with high figure of merit\cite{Shafique2017}, especially for the crystal SnSe with an extraordinarily high thermoelectric zT value of  of $2.6 \pm 0.3$ at 923 K\cite{Zhao2014}, generally possessing puckered structures and behaving like semiconducting, is believed to be a family of ``phosphorene analogues''\cite{Gomes2015}.  Rohr  \textit{et al.} predicted recently a new polymorph of GeSe\cite{Rohr2017}, called as $\beta-$GeSe, which is a boat conformation for its Ge-Se six-membered ring. Experimental measurement and theoretical calculations reveal that $\beta-$GeSe, similar to another polymorph of GeSe, i.e. $\alpha-$GeSe\cite{Wiedemeier1978}, is a semiconductor with a moderate band gap, which make it promising for future electronic and optoelectronic applications. Further investigations on monolayer $\beta-$GeSe and the related polymorph are thus necessary to gain insights on this new kind of 2D materials.

In this work, we systematically investigate the electronic, transport and optical properties of monolayer $\alpha-$ and $\beta-$GeSe by using first-principles calculations. We demonstrate that the direct band gap of monolayer $\alpha-$GeSe is smaller than the indirect band gap of $\beta-$GeSe. Both monolayers of $\alpha-$ and $\beta-$ GeSe have exceptionally high electron mobility, which we predict to be 4.24 $\times10^3$$ cm^2/V\cdot {s}$ and 7.84 $\times10^4$$ cm^2/V\cdot {s}$ respectively with strong anisotropy. Furthermore, we also investigate the strain-engineering and optical properties of these two materials.

\section{Method and computational details}
The calculations are performed using the Vienna \textit{ab-initio} simulation package (VASP) based on density functional theory \cite{Kresse1996}.  The exchange-correlation energy is described by the generalized gradient approximation (GGA) using the Perdew-Burke-Ernzerhof (PBE) functional. The calculation is carried out by using the projector-augmented-wave pseudopotential method with a plane wave basis set with a kinetic energy cutoff of 600 eV. When optimizing atomic positions, the energy convergence value between two consecutive steps is chosen as 10$^{-5}$ eV and the maximum Hellmann-Feynman force acting on each atom is 10$^{-3}$ eV/\AA. For $\alpha-$GeSe monolayer, the Monkhorst-Pack scheme is used for the Brillouin zone integration with k-point meshes of 17$\times $15$\times $1 and 25$\times $21$\times $1 for geometry optimization and self-consistent electronic structure calculations, respectively. And for $\beta$ phase, we use 17$\times $11$\times $1 and 25$\times $15$\times $1 Monkhorst-Pack $k$ meshes for the structure relaxation and electronic structure calculations, respectively. To verify the results of the PBE calculations, the electronic structures of $\alpha-$ and  $\beta-$GeSe are calculated using the much more computationally expensive hybrid Heyd-Scuseria-Ernzerhof (HSE06) functional\cite{HSE03,HSE06}. Generally, HSE06 improves the precision of band gap by reducing the localization and delocalization errors of PBE and HF functionals. Hereby, the HSE06 calculations incorporate 25\% short-range Hartree-Fock exchange. The screening parameter u is set to 0.4 \AA$^{-1}$. The complex dielectric functions $\epsilon(\omega)$ of monolayer $\alpha-$ and $\beta-$GeSe are calculated by using HSE06 hybrid functional on a grid of 11$\times $11$\times $1.

The optical properties of monolayer $\alpha-$ and $\beta-$GeSe are obtained based on the results of complex dielectric function, $i.e.$ $\epsilon(\omega)=\epsilon_1(\omega)+i\epsilon_2(\omega)$. The imaginary part of dielectric tensor $\epsilon_2(\omega)$ is determined by a summation over empty band states as follows \cite{Gajdos2006},

\begin{equation}
\epsilon_2(\omega) = \frac{2\pi e^2}{\Omega \epsilon_0} \sum_{k,v,c} \delta(E_k^c-E_k^v-\hbar \omega) \Bigg\vert\langle \Psi_k^c \big\vert \textbf{u}\cdot\textbf{r} \big\vert \Psi_k^v \rangle \Bigg\vert ^2,
\end{equation}

where $\epsilon_0$ is the vacuum dielectric constant, $\Omega$ is the crystal volume, $v$ and $c$ represent the valence and conduction bands respectively, $\hbar\omega$ is the energy of the incident photon, \textbf{u} is the vector defining the polarization of the incident electric field, \textbf{u}$\cdot$\textbf{r} is the momentum operator, $\Psi_k^c$ and $\Psi_k^v$ are the wave functions of the conduction and valence band at the $k$ point, respectively. The real part of dielectric tensor $\epsilon_1(\omega)$ is obtained by the well-known Kramers-Kronig relation\cite{dresselhaus1999solid},

\begin{equation}
\epsilon_1(\omega)=1+\frac{2}{\pi}P\int_0^{\infty} \frac{\epsilon_2(\omega ')\omega '}{\omega '^2-\omega^2+i\eta}d\omega ',
\end{equation}

where $P$ denotes the principle value. The absorption coefficient $\alpha(\omega)$ and reflectivity $R(\omega)$ can be subsequently given by \cite{Saha2000,Luo2015,xu2017}

\begin{equation}
\alpha(\omega)=\frac{\sqrt{2}\omega}{c} \Big\lbrace \big[\epsilon_1^2(\omega)+\epsilon_2^2(\omega)\big]^{1/2}-\epsilon_1(\omega) \Big\rbrace ^{\frac{1}{2}},
\end{equation}

\begin{equation}
R(\omega)=\Bigg| \frac{\sqrt{\epsilon_1(\omega)+i\epsilon_2(\omega)}-1}{\sqrt{\epsilon_1(\omega)+i\epsilon_2(\omega)}+1} \Bigg| ^2,
\end{equation}

By using the deformation potential theory for semiconductors, which was proposed by Bardeen and Shockley\cite{deformation1950}, the intrinsic carrier mobility $\mu$ of monolayer $\alpha-$ and $\beta-$GeSe is calculated and investigated in details herein. In the long-wavelength limit, when only considering the interaction between electron and longitudinal acoustic phonon\cite{deformation1950}, the carrier mobility of 2D semiconductors is given by \cite{Cai2014, Shuai2011, Shuai2013, Wang2015b},

\begin{equation}\label{mobilities}
\mu = \frac{2e\hbar^3C}{3k_BT|m^*| ^2E_l^2},
\end{equation}

where $e$ is the electron charge, $T$ is the temperature equal to 300 K throughout the paper. $C$ is the elastic modulus of a uniformly deformed crystal by strains and derived from ${C^{2D}_{\alpha}}=[{\partial^2{E}/\partial^2(\delta{l}/l_0)}]/{S_0}$, while $E$ is the total energy, $\delta{l}$ is the change of lattice constant $l_0$ along the transport direction $\alpha$, and $S_0$ represents the lattice volume at equilibrium for a 2D system, $m^*$ is the effective mass given by $m^*=\hbar^2({\partial^2{E(k)}/{\partial{k^2}})}^{-1}$ ($\hbar$ is the reduced Planck's constant, $k$ is wave-vector, and $E(k)$ denotes the energy). $E_l$ is the deformation potential (DP) constant defined by $E_l^{e(h)}=\Delta{E_{CBM(VBM)}}/(\delta{l}/l_0)$, where $\Delta{E_{CBM(VBM)}}$ is the energy shift of the band edge with respect to the vacuum level under a small dilation $\delta{l}$ of the lattice constant $l_0$. 

\section{Results and discussion}
\subsection{Geometric structure of monolayer $\alpha-$ and $\beta-$GeSe}
\begin{figure}
\centering
\includegraphics[width=0.8\linewidth]{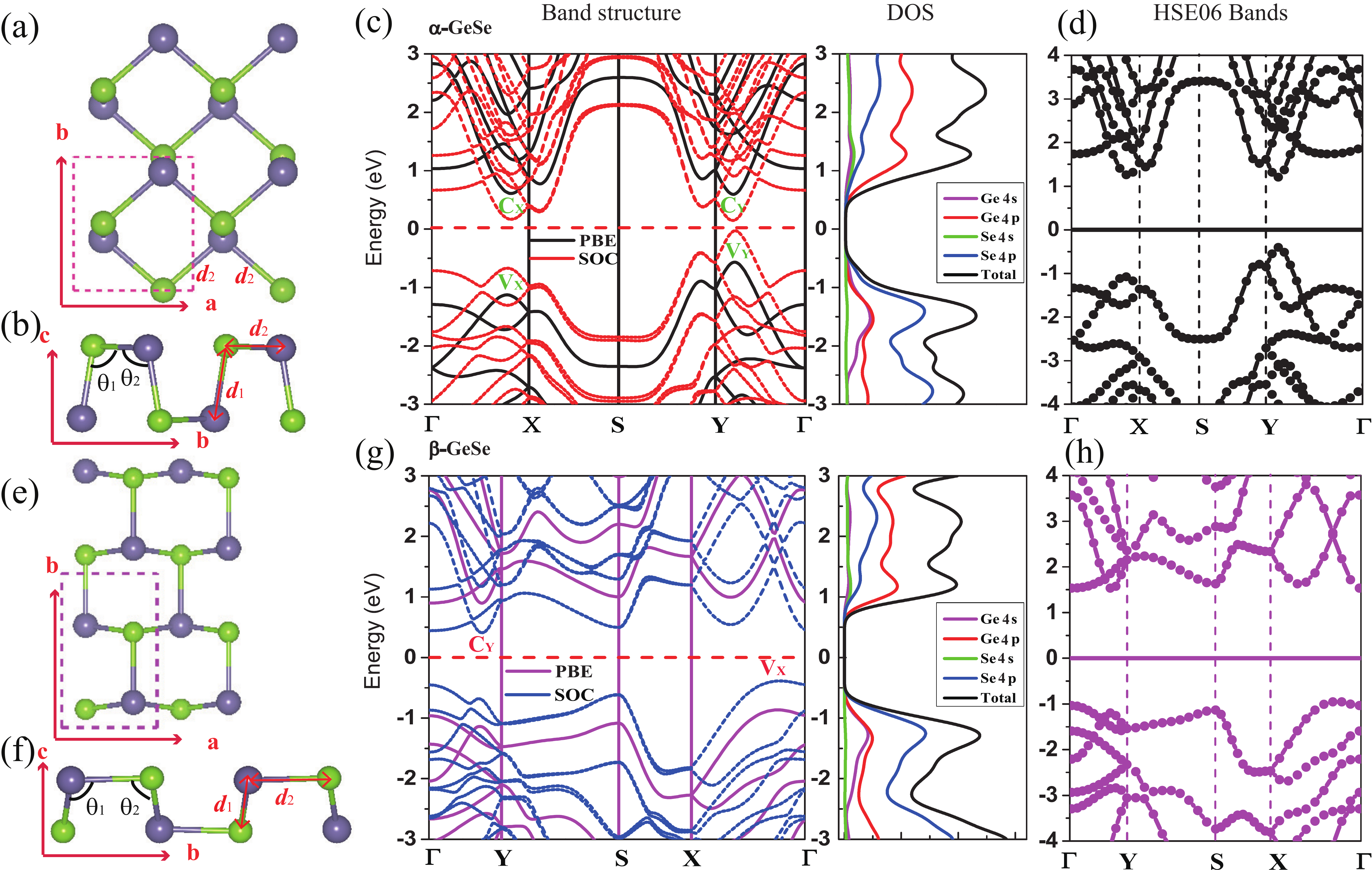}
\caption{Atomic structure of  monolayer $\alpha-$ and $\beta-$GeSe in a 2$\times $2$\times $1 supercell from top view (a,b) and side view (e,f), respectively. The $a/b$ direction is the armchair/zigzag direction, respectively. The blue and green balls denote Ge and Se atoms respectively. (c) and (g) are PBE calculations with and without SOC of electronic band structure and PBE calculated density of states (DOS) of monolayer $\alpha-$GeSe and $\beta-$GeSe along high-symmetry directions respectively. (d) and (h) are the electronic band structures under HSE06 hybrid functional.}
\label{structure-band} 
\end{figure}

\begin{table}
\centering
\caption{Structural informations, SOC strength and band gaps of monolayer $\alpha-$ and $\beta-$GeSe. The bond angles of $\theta$ and bond lengths of $d$ are indicated in Fig. \ref{structure-band}.}
\begin{tabular}{cccccccccc}
\hline
  phase & $a_0$ &  $b_0$ &  $d_1$&  $d_2$[\AA] &  $\theta_1$  &  $\theta_2$[$^{\circ}$] &$E^{SOC}_{Ge}$ &$E^{SOC}_{Ge}$& $E_g(PBE/SOC/HSE06)[eV]$  \\
\hline
$\alpha$ & 3.97& 4.29 &2.54 & 2.66 &96.59 & 97.41 &-0.016&-0.040&1.16/0.18/1.61 \\
$\beta$ & 3.67 & 5.91 &2.55 &2.72 &96.65 & 93.91&-0.016&-0.041&1.76/0.80/2.47 \\
\hline
\end{tabular}
\label{tabel-structure}
\end{table}

The top and side view of the fully optimized structure of monolayer $\alpha-$ and $\beta-$GeSe with space group of Pmn2-1(31) are shown in Fig.~\ref{structure-band} ((a), (b)) and ((e), (f)), respectively. The tetragonal unit cell (the dashed rectangle) of monolayer $\alpha-$ and $\beta-$GeSe contains two germanium atoms and two selenium atoms. From the top view of $\alpha-$ and $\beta-$GeSe as shown in Fig.~\ref{structure-band} (a) and (e), monolayer $\alpha-$GeSe is a puckered structure analogue to black phosphorene with Ge and S atoms substituted for P atoms alternately\cite{Feng2016, Qiao2014, Fei2014} and $\beta-$GeSe consists of six-rings with the vertices arranged in an uncommon boat conformation\cite{Fabian2017}, as shown in the side view in Fig.~\ref{structure-band} (f). From the side views in Fig.~\ref{structure-band} (b,f), both monolayer $\alpha-$ and $\beta-$GeSe consist of two atomic sublayers. The $\alpha-$ and $\beta-$GeSe monolayer are optimized with vacuum layers of 21 $\AA$ and 16$\AA$, respectively, which are large enough to avoid the artificial interaction between atom layers. 

Based on the first-principles method, the lattice constants of $\alpha-$GeSe were calculated to be $a_0$ = 3.97 $\AA$ and $b_0$ = 4.29 $\AA$, which are in good accordance with previous theoretical and experimental results\cite{Fabian2017, Hu2015, Fei2015, Shengli2015}. Each Ge/Se atom binds three neighboring Se/Ge atoms with different bond lengths, as shown in Figs. \ref{structure-band}(a,b), and two of them are identical with $d_2$ = 2.66 $\AA$ and the other one is $d_1$ = 2.54 $\AA$. The bond angles of $\theta_{GeSeGe}$ and $\theta_{SeGeSe}$ are 96.59$^{\circ}$ and  97.41$^{\circ}$, respectively. For monolayer $\beta-$GeSe, the optimizied lattice constants are $a_0$ = 3.67 $\AA$ and $b_0$ = 5.91 $\AA$, which are in good agreement with previous experimental results\cite{Fabian2017}. In this boat structure, each Ge atom binds three Se atoms with two identical bond lengths of $d_1$ = 2.55 $\AA$ and one bond length of $d_2$ = 2.72 $\AA$, and the bond angles of $\theta_{GeSeGe}$ and $\theta_{SeGeSe}$ are 93.91$^{\circ}$ and  96.65$^{\circ}$, respectively. 

\subsection{ Electronic band structure and stability of 2D $\alpha-$ and $\beta-$GeSe}
The calculated electronic band structures performed by both PBE (with and without considering spin-orbital-coupling, SOC) and HSE06-hybrid functional method for monolayer $\alpha-$ and $\beta-$GeSe along high-symmetry directions of Brillouin zone (BZ) are shown in Fig.~\ref{structure-band} (c-d) and (g-h) respectively. 

For monolayer $\alpha-$GeSe shown in Fig.~\ref{structure-band} (c), both the conduction band minimum (CBM) and the valence band maximum (VBM) are located along the $\Gamma-Y$ direction denoted by $C_y$ and $V_y$ respectively. The formation of the band gap of $\alpha-$GeSe shows that it is a semiconductor with a direct band gap of 1.16 eV, which is well consistent with the previous reported results\cite{Hu2015, Shafique2017}. However, it is worthy to mention that the energy difference (0.014 eV) between the local CBMs ($C_X$ and $C_Y$ in Fig.~\ref{structure-band} (c)) and the CBM at $k_{CBM}$ is very small, and such nearly degenerate VBMs suggest that it is feasible to apply some kinds of external controls (e.g. strain), to tune monolayer $\alpha-$GeSe from direct to indirect semiconductors, or vice versa. Since Ge and Se are relatively heavy elements, the SOC effect may influence the band structures obviously, which is confirmed by the calculated band structures with SOC involved, as shown as in Fig.~\ref{structure-band} (c) (the red dashed line). The corresponding HSE06 calculation in Fig.~\ref{structure-band} (d) gives a larger band gap of 1.61 eV compared to the PBE result of 1.16 eV, since PBE always underestimates the value of band gaps of semiconductors.

Fig.~\ref{structure-band} (g) shows the PBE calculations with and without SOC for the electronic band structure and DOS of monolayer $\beta-$GeSe, which shows that, monolayer $\beta-$GeSe is a semiconductor with an indirect band gap with $E_g$ = 1.76 eV. The CBM locates at the midpoint along $\Gamma-Y$ direction and the VBM locates near the $\Gamma$ point along the $\Gamma-X$ direction. The obtained band gap decreases to 0.80 eV when the SOC is involved. The HSE06 calculation without SOC effects gives the band gap of 2.47 eV, which is larger than that of $\alpha-$GeSe. 

As for the SOC effects, since the calculated SOC strength is negative as shown in Table.~\ref{tabel-structure}, thus the calculated band gaps for both $\alpha-$ and $\beta-$GeSe decrease when SOC is involved.

\begin{figure}
\centering
\includegraphics[width=1\linewidth]{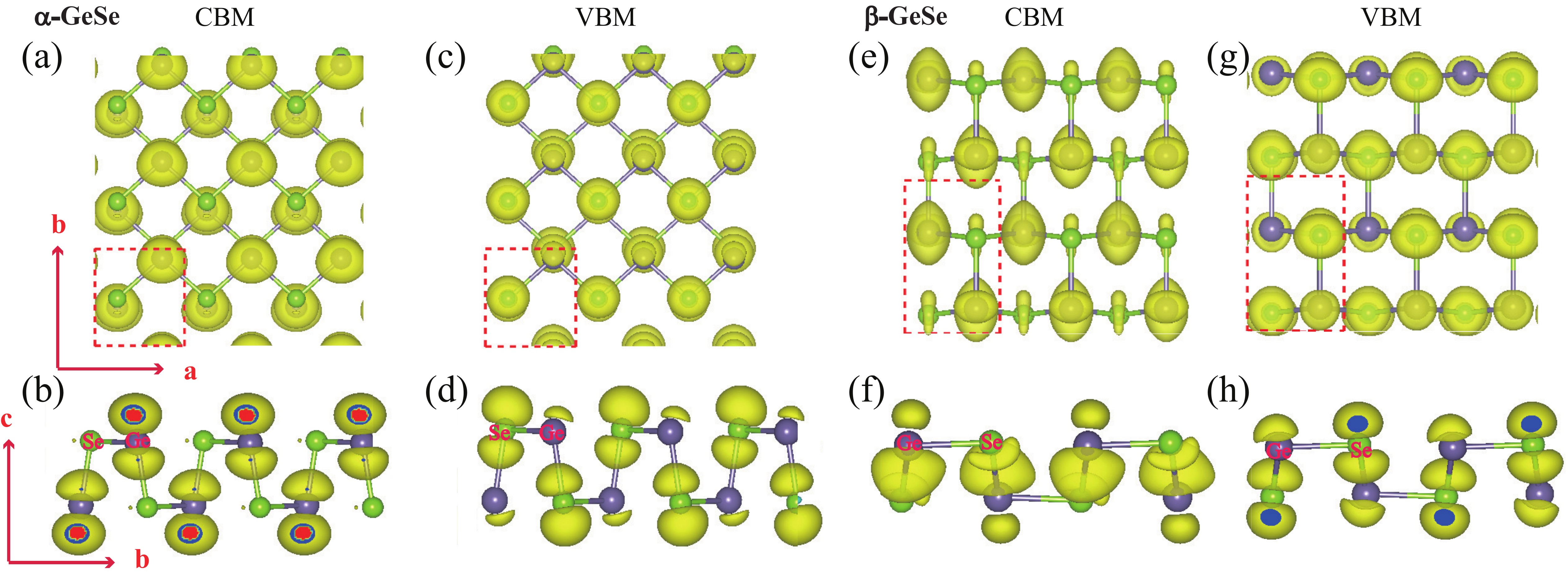}
\caption{Iso-surface plots of the charge density of CBM ((a, b)), VBM ((c, d)) for monolayer $\alpha-$GeSe and CBM ((e, f)), VBM ((g, h)) for monolayer $\beta-$GeSe illustrated in the $ab$ and  $bc$ plane, with an iso-value of of 0.007 $e/\AA^3$.}
\label{charge} 
\end{figure}

In order to clarify the contributions from different orbitals to the band structures around the Fermi level of $\alpha-$ and $\beta-$GeSe, we calculate the total and partial density of states (DOS) as shown in the right part of  Fig.~\ref{structure-band} (c) and (g). Analysis on the PDOS (Ge-4$s$, 4$p$ and Se-4$s$, 4$p$ orbitals) of $\alpha-$GeSe reveals that Ge-4$p$ and Se-4$p$ orbitals dominate the electronic states near the Fermi level. The contributions from the Ge-4$p$ to the total DOS of the conduction bands is larger than that from Se-4$p$, while in the valence band, the Se-4$p$ orbitals contribute larger than that from Ge-4$p$. Analysis on the the PDOS of monolayer $\beta-$GeSe reveals the dominant contributions from Ge-4$p$ and Se-4$p$ orbitals to the total DOS near the Fermi level, and the respective contribution from the Ge-4$p$ and Se-4$p$ orbitals is similar to the case of $\alpha-$GeSe.

The above-mentioned PDOS analysis on the orbitals contributions to the formation of CBM and VBM is validated by the partial charge densities associated with the CBM and VBM, as shown in Fig.~\ref{charge}. In both materials, the p orbitals of Ge atoms contributes dominantly to CBM by connecting the neighboring Se atoms via bonding states, however, VBM is dominated by the contribution from p orbitals of Se atoms by connecting the neighboring Ge atoms via antibonding states.

\subsection{Strain-engineering electronic properties of monolayer $\alpha-$ and $\beta-$GeSe}

\begin{figure}
\centering
\includegraphics[width=0.8\linewidth]{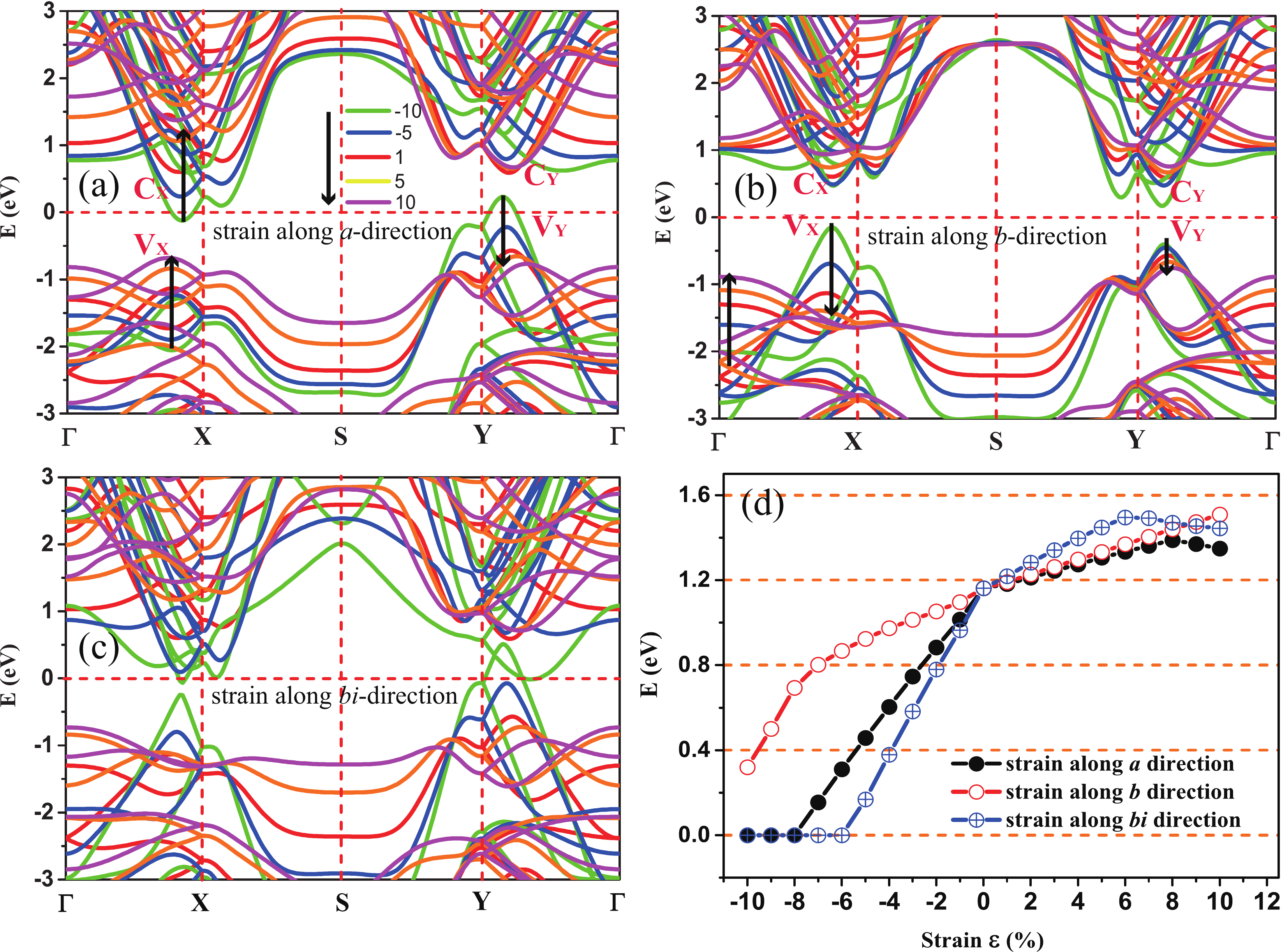}
\caption{PBE calculated band structures under uniaxial ((a) $a$ direction,(b) $b$ direction) and biaxial (c) strain of monolayer $\alpha-$GeSe. (d) shows the evolution of band gaps for $\alpha-$GeSe as a function of the applied strain.}
\label{a-strain} 
\end{figure}

\begin{figure}
\centering
\includegraphics[width=0.8\linewidth]{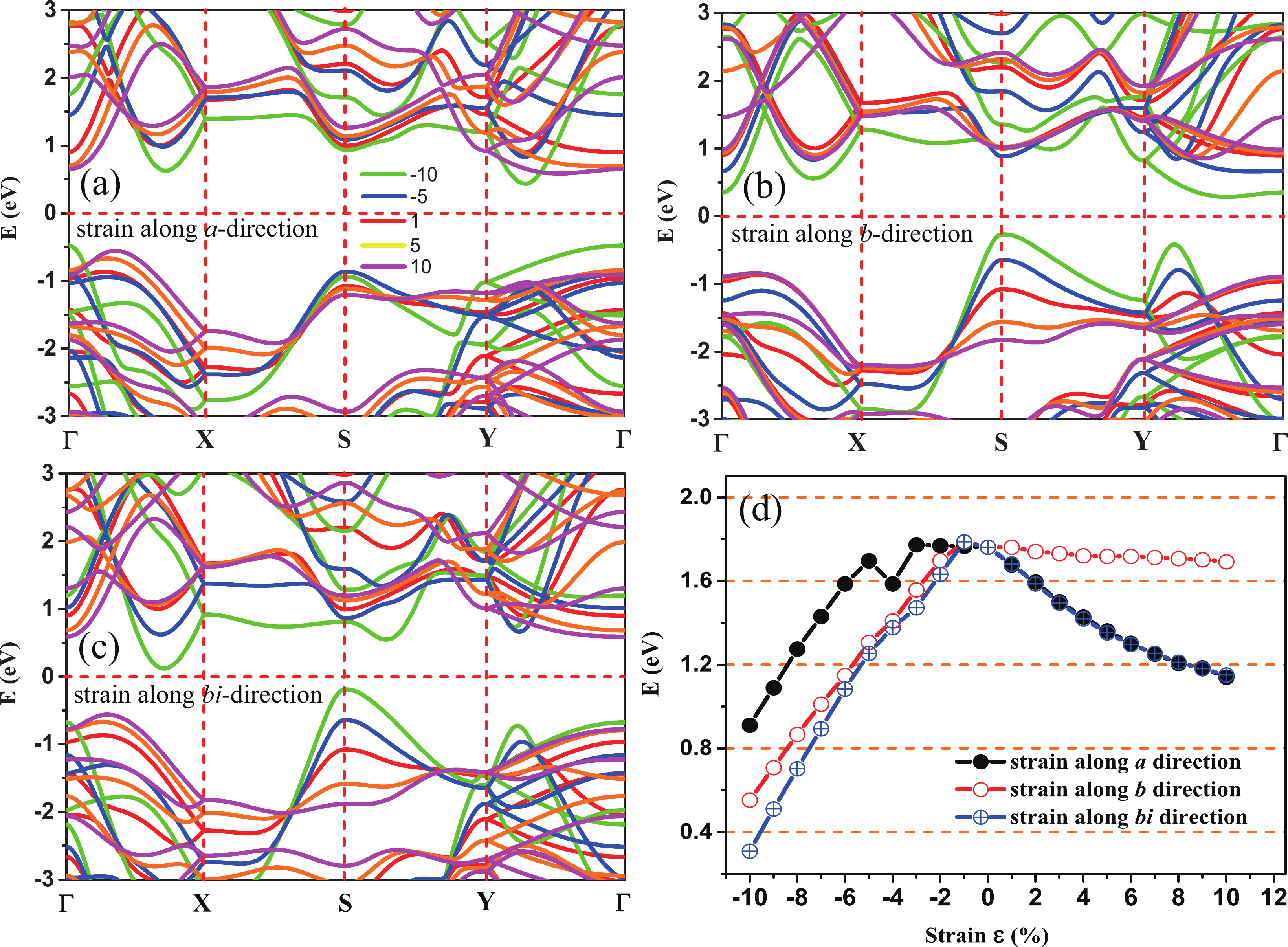}
\caption{PBE calculated band structures under uniaxial ((a) $a$ direction,(b) $b$ direction) and biaxial (c) strain of monolayer $\beta-$GeSe. (d) shows the evolution of band gaps for $\alpha-$GeSe as a function of the applied strain.}
\label{b-strain} 
\end{figure}

Here, we studied the effects of in-plane uniaxial ($a$ and $b$ direction) and biaxial ($bi$: $a$ + $b$ direction) strains on electronic properties of monolayer $\alpha-$ and $\beta-$GeSe, to realize the possible tunability of their electronic properties.  In this work, $\varepsilon_x$, $\varepsilon_y$ and $\varepsilon_{xy}$ indicate the components of the relative strain along $a$, $b$ and $bi$ directions respectively. The positive (negative) values represent tensile (compressive) strain, and evaluated as the lattice stretching (condensing) percentage.

 Fig.~\ref{a-strain} shows the valence and conduction band structures of monolayer $\alpha-$GeSe as a function of different strains, from -10\% to 10\% of the fully relaxed structure. Fig.~\ref{a-strain} (a-c) show the dependence of the energy bands on strain along $a$, $b$ and $bi$ direction, respectively. Fig.~\ref{a-strain} (d) shows the evolution of calculated band gap under various strains. 

The VBMs ($V_X$ and $V_Y$) and CBMs ($V_X$ and $V_Y$) are shown in Fig.~\ref{a-strain}. Monolayer $\alpha-$GeSe without strains is a direct semiconductor described by VBM of $V_Y$ and CBM of $C_Y$. When compressive strains are applied along $a$ direction, the values of $V_Y$ and $C_Y$ increase, while the value of $C_X$ decreases, leading to a transition from direct to indirect semiconductor of monolayer $\alpha-$GeSe even at a small compression with $\varepsilon_x$=-1\%. When a compressive strain of $\varepsilon_x$=-8\% is applied, a semiconductor-to-metal transition takes place as shown in Fig.~\ref{a-strain} (d). 

However, when applying a tensile strain along $a$ direction, the changes of the local VBM and CBM are different. When the tensile strain increases, the value of local $V_Y$ decreases, and the values of $C_X$ and $V_X$ increase, while the value of $C_Y$ keeps nearly unchanged, subsequently leading to a transition from direct to indirect semiconductors at $\varepsilon_x$=9\%. So, the direct-semiconducting characteristic (from $\varepsilon_x$=-1\% to $\varepsilon_x$=9\%) of monolayer $\alpha-$GeSe seems robust for a moderate stretching.

The change of the band structures of monolayer $\alpha-$GeSe when applying an external strain along $b$ direction ($\varepsilon_y$) is much smaller than that for strains along $a$ direction, similar analysis on the evolution of band structures for compressive and tensile strains, as shown in Figs.~\ref{a-strain} (b) and (d), show that, for applied strains from -10\% to 10\%, a transition from direct to indirect semiconductor happens, while the transition from semiconductor to metal does not takes place in the strain region.

When applying strains along the $bi$ direction from -10\% to 10\%, the changes of the band structures are more similar to those for strains along the $a$ direction, and the transitions for both direct-to-indirect semiconductor and semiconductor-to-metal happen,  as shown in Figs.~\ref{a-strain} (c) and (d), except that the latter transition occurs at a compressive strain of $\varepsilon_{xy}$=-6\%.

In addition, the evolution of the values of band gaps of $\alpha-$GeSe under various strains shown in Fig.~\ref{a-strain} (d) indicates that the values of the calculated band gaps increase with the increase of tensile strains in the region from -10\% to 10\%, disregarding the directions the strains apply.

Fig.~\ref{b-strain} shows the evolution of band structures and values of band gap of monolayer $\beta-$GeSe as a function of compressive and tensile strains with the strength from -10\% to 10\%. The behavior of $\beta-$GeSe under strains is quite different compared to that of $\alpha-$GeSe, obviously without the above-mentioned transition from semiconductor to metal in this strain region. Furthermore, although the band gap of monolayer $\beta-$GeSe decreases when the applied compressive strains increase, which is similar to the case of $\alpha-$GeSe, however, the tendency of band gap is opposite to that of $\alpha-$GeSe when applying tensile strains. More interestingly, the calculated band gaps under tensile strain along $b$ direction keep nearly constant when the tensile strain increase in the region from 0\% to 10\%, and the value of band gap is around 1.73 eV. Such kind of robust band gap nearly disregarding the tensile strains subsequently lead to the nearly overlap of band gaps for the two cases by applying tensile strains along $a$ and $bi$ directions, as shown in Fig.~\ref{b-strain}(d).

\subsection{Transport properties of monolayer $\alpha-$ and $\beta-$GeSe}
\begin{figure}
\centering
\includegraphics[width=0.8\linewidth]{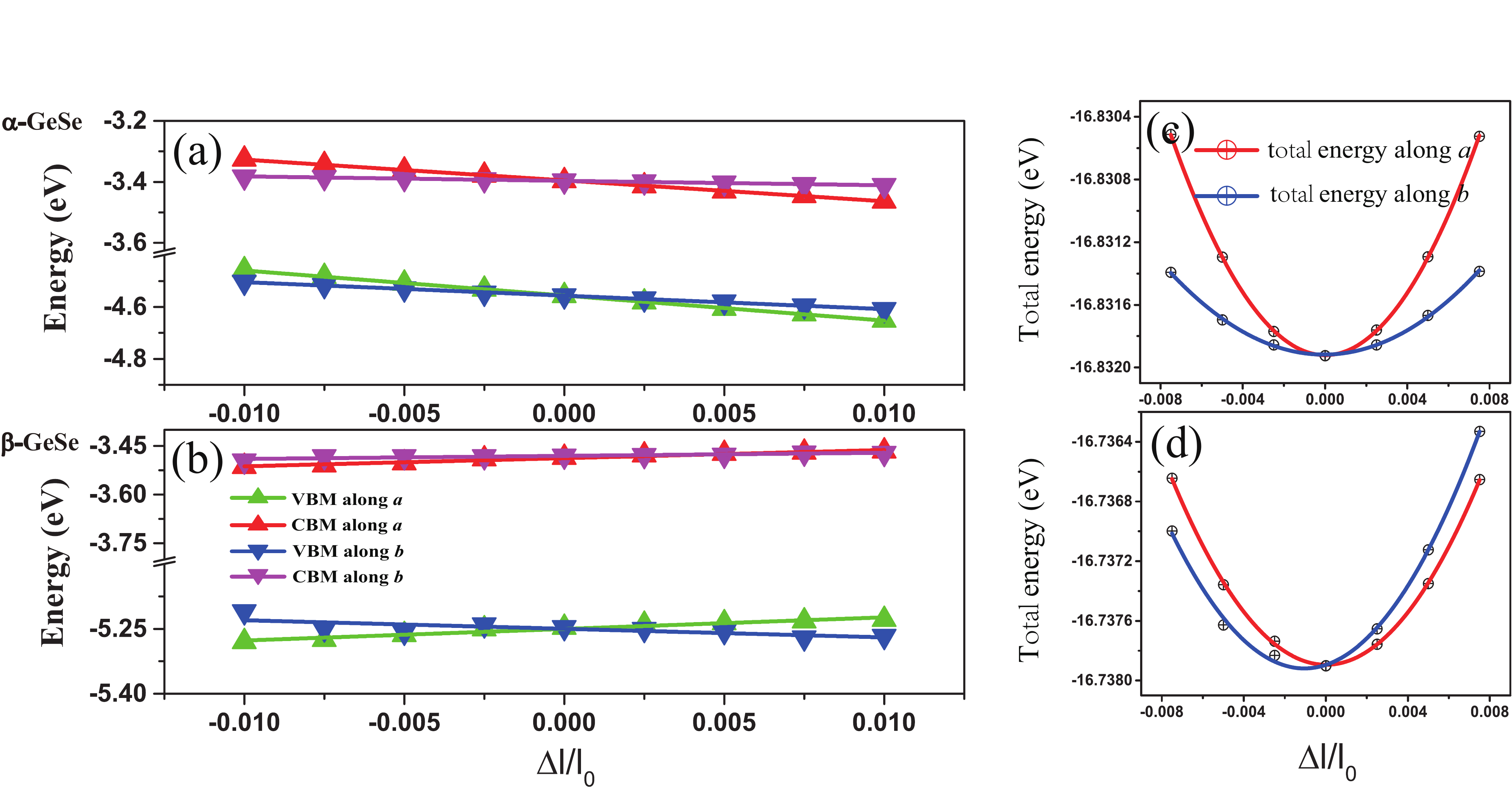}
\caption{Dependence of band edges with respect to vacuum as a function of applied uniaxial strains along $a$ and $b$ directions for monoalyer (a) $\alpha-$ and (b) $\beta-$GeSe, respectively. (c) and (d) shows the relationship between total energy and strain along $a$ and $b$ directions for $\alpha-$ and $\beta-$GeSe, respectively.}
\label{deformation} 
\end{figure}
To investigate the electronic transport properties of 2D GeSe and their potential for electronic applications, we calculate the carrier mobilities of monolayer $\alpha-$ and $\beta-$GeSe. According to the deformation theory, i.e. Eq. (5), three parameters, namely carrier effective mass $m^*$, the deformation potential $E_l$ and the elastic modulus $C_{2D}$ in the propagation direction determine the behaviors of carrier mobility of semiconductors\cite{Shuai2012, Shuai2013}. Although the PBE calculations always underestimate the band gap, the curvatures of valence and conduction bands calculated by PBE method are generally correct enough and the calculated carrier mobilities are subsequently in good agreement with experiments for 2D materials\cite{Dai2015, Cai2014, Shuai2013, Shuai2012, Shuai2011}. 
\begin{table}
\centering
\caption{Calculated effective mass $m^*$ (with $m_0^*$ being the static electron mass), deformation potential constant $E_{l}$, 2D elastic modulus $C$ and carrier mobility $\mu$ along the $\Gamma-Y$ direction. The electron and hole carrier mobilities $\mu$ are calculated by using Eq.(1) at $T$=300 K.}
\begin{tabular}{ccccccccccccc}
\hline
  Phase&carrier type&  $m_a^*$ &  $m_b^*$ &  $E_{l-a}$ &  $E_{l-b}$&  $C_a$  &  $C_b$  & $\mu_a$ & $\mu_b$\\
 &&[$m_0$]&  & [$eV$] &  &  [$N/m$] & & [$ cm^2/V\cdot {s}$] &[$ cm^2/V\cdot {s}$] \\
\hline
$\alpha$&hole& 0.33& 0.16 &9.67&5.20&37 & 14 &50 & 3.02$\times10^2$ \\
          &electron&  0.27 &0.15&6.83&1.45&37 & 14 & 1.52 $\times10^2$&4.24$\times10^3$ \\
$\beta$ &hole &1.09 &0.84&2.68&1.99&42 & 41 & 70 &2.06$\times10^2$ \\
& electron & 1.45 &0.09&2.52&0.96& 42 & 41& 45& 7.84$\times10^4$ \\
\hline
\end{tabular}
\label{mobility}
\end{table}

Firstly, we calculate the effective masses of holes ($m_h^*$) and electrons ($m_e^*$) along $a$ and $b$ directions by fitting parabolic functions to the band close to the VBM and CBM, respectively, as shown in Table.~\ref{mobility}. For monolayer $\alpha-$GeSe, the hole effective mass ($m_h^*$) along $a$ direction (0.33 $m_0$) is larger than that along $b$ direction (0.16 $m_0$), but the difference of electron effective mass ($m_e^*$) between $a$ and $b$ directions is trivial. Contrarily, the electron effective mass of monolayer $\beta-$GeSe displays a strong anisotropy (1.45 $m_0$ along $a$ direction and 0.09 $m_0$ along $b$ direction).

Then, through stretching and compressing the lattices along $a$ and $b$ directions respectively, the elastic constants and the DP constants can be calculated. By linearly fitting the band energy (CBM and VBM positions) shift with respect to the vacuum level under strain $\varepsilon$ along $a$ and $b$ directions, we obtain the DP constant $E_l$ for both electrons and holes. As shown in Fig.~\ref{deformation} (a) and (b),  the response to of CBM and VBM to the applied strain appears to be highly anisotropic. The obtained DP constants are listed in Table.~\ref{mobility} as well. In Fig.~\ref{deformation} (c) and (d) we present the total energy in monolayer $\alpha-$ and $\beta-$GeSe as a function of the lattice dilation along $a$ and $b$ direction, and all the calculated elastic constant are listed in Table.~\ref{mobility}.

Based on the obtained $m^*$, $E_l$ and $C^{2D}$, the carrier mobilities of 2D $\alpha-$ and $\beta-$GeSe at room temperature ($T$=300 K) are calculated and listed in Table.~\ref{mobility}. The predicted carrier mobilities for both $\alpha-$ and $\beta-$GeSe are anisotropic along the $a$ and $b$ directions. Both $\alpha-$ and $\beta-$GeSe show high electron mobilities along the $b$ direction ($\mu_b$), i.e. 4.24 $\times10^3$$ cm^2/V\cdot{s}$ for monolayer $\alpha-$GeSe and 7.84 $\times10^4$$ cm^2/V\cdot{s}$ for monolayer $\beta-$GeSe, respectively. The calculated value of $\mu_b$ for $\beta-$GeSe is even larger than that of black phosphorene (1.10$\sim$1.14 $\times10^4$$ cm^2/V\cdot{s}$)\cite{Qiao2014}, which means that monolayer $\beta-$GeSe will be a promising candicate material for future electronic and optoelectronic applications, considering the high electron mobilities and stability in ordinary environment\cite{Rohr2017}. 

\subsection{Optical properties of monolayer $\alpha-$ and $\beta-$GeSe}
\begin{figure}
\centering
\includegraphics[width=0.75\linewidth]{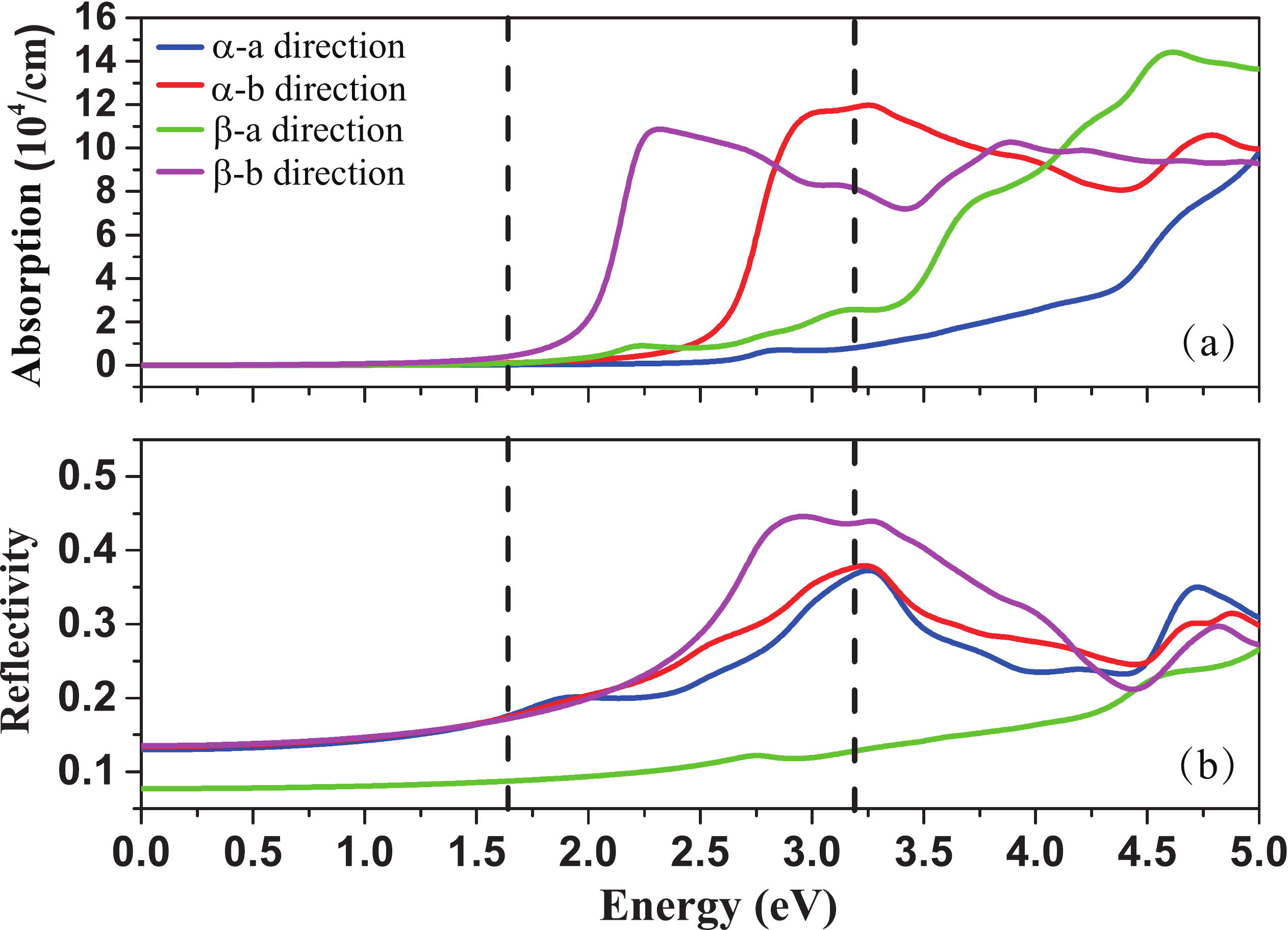}
\caption{HSE06 calculations of (a) optical absorption spectra and (b) reflectivity of monolayer $\alpha-$ and $\beta-$GeSe for incident light with the polarization along the $a$ and $b$ directions.}
\label{optical} 
\end{figure}
Fig.~\ref{optical} presents the  in-plane optical absorption spectra $\alpha(\omega)$ and reflectivity $R(\omega)$ of monolayer $\alpha-$ and $\beta-$GeSe for the incident light with the polarization of the electric field E along the $a$ ($\textbf{E}//a$) and $b$ ($\textbf{E}//b$) directions respectively. The corresponding real and imaginary parts of the dielectric function are shown in Fig. S1 in the supplementary information. 

The absorption coefficient is defined as the decay of light intensity spreading in a unit length, i.e. Eq. (3). For monolayer $\alpha-$ and $\beta-$GeSe, the absorption coefficient shows an obvious anisotropy along $a$ and $b$ directions, and both materials exhibit obvious optical absorption with the frequency covering some part of the visible spectrum. The frequency region for high absorption along the $b$ direction of $\beta-$GeSe is larger than that along the $a$ direction shown in Fig.~\ref{optical} (a), and is even larger than those of $\alpha-$GeSe irrespective of polarization directions. Such a significant anisotropic optical property can be used to identify the monolayer $\beta-$GeSe in experiments.

Fig.~\ref{optical} (b) shows the reflectivity $R(\omega)$  for both monolayer $\alpha-$ and $\beta-$GeSe. The isotropic property of $\alpha-$GeSe and anisotropic for $\beta-$GeSe are observed in the reflectivity curves as shown in Fig.~\ref{optical} (b). For $\beta-$GeSe, $R(\omega)$ along the $b$ direction in the visible region is higher than that of $\alpha-$GeSe along both $a$ and $b$ directions, which thus means that monolayer $\beta-$GeSe is polarizationally non-transparent material.

\section{Conclusion}
In summary, we have performed first-principles calculations on the structure, electronic, charge carrier transport and optical properties of monolayer $\alpha-$ and $\beta-$GeSe. $\alpha-$GeSe is a puckered structure similar to that of black phosphorene with Ge and S atoms substituted for P atoms alternately, while $\beta-$GeSe consists of six-rings with the vertices arranged in an uncommon boat conformation. The all have strongly anisotropic properties. For monolayer $\alpha-$GeSe, the direct-semiconducting characteristic (from $\varepsilon_x$=-1\% to $\varepsilon_x$=9\%) seems robust for a moderate stretching. For $\beta-$GeSe, along the zigzag direction, it has an extremely high electron mobility of 7.84 $\times10^4$$cm^2/V\cdot {s}$ and the calculated band gaps keep nearly unchanged under tensile strain from 0\% to 10\%. The calculated optical properties of monolayer $\alpha-$ and $\beta-$GeSe shows anisotropic behaviors and large absorption in some parts of visible spectrum. By comparison, monolayer $\beta-$GeSe is predicted to be promising in the future electronic and optoelectronic applications due to the ultrahigh electron mobilities and the abnormal behavior of robust band gap disregarding the applied tensile strains.

\section*{Acknowledgement}
This work is supported by the National Natural Science Foundation of China under Grants No. 11374063 and 11404348,nd the National Basic Research Program of China (973 Program) under Grants No. 2013CBA01505.

\section*{Reference}
\bibliographystyle{unsrt}
\bibliography{GeSe}

\end{document}